\documentclass[10pt,conference]{IEEEtran}
\usepackage[utf8]{inputenc}

\usepackage{soul}
\usepackage{xcolor}
\usepackage{xspace}
\usepackage{url}
\usepackage{graphicx}
\usepackage{listings,multicol}
\usepackage[export]{adjustbox}
\usepackage{amsmath}
\usepackage{amssymb}
\usepackage{textcomp}
\usepackage{breakurl} 
\usepackage{listings}
\usepackage{algorithm}
\usepackage{hyperref}
\usepackage{cleveref}
\usepackage{subcaption}
\usepackage[noend]{algpseudocode}

\title{The IntelliJ Platform: a Framework for Building Plugins and Mining Software Data}

\author{
 \IEEEauthorblockN{Zarina Kurbatova,\IEEEauthorrefmark{1} Yaroslav Golubev,\IEEEauthorrefmark{1} Vladimir Kovalenko,\IEEEauthorrefmark{1}\IEEEauthorrefmark{2} Timofey Bryksin\IEEEauthorrefmark{1}\IEEEauthorrefmark{4}}
    \IEEEauthorblockA{\IEEEauthorrefmark{1}\textit{JetBrains Research},  \IEEEauthorrefmark{2}\textit{JetBrains N.V.}, \IEEEauthorrefmark{4}\textit{Saint Petersburg State University}}
    \IEEEauthorblockA{\{zarina.kurbatova, yaroslav.golubev, vladimir.kovalenko,  timofey.bryksin\}@jetbrains.com}
}

\begin{document}

\maketitle

\begin{abstract}
    In software engineering, a great number of new approaches are being actively researched, and a lot of tools are being developed based on them. These tools require a framework for their creation and an opportunity to be used by potential developers. Modern IDEs provide both.
    
    In this paper, we describe the main capabilities of the IntelliJ Platform that could be useful for researchers that are developing code analysis tools. To illustrate the benefits of using the platform, we describe several use cases that researchers might be interested in: mining software data, running machine learning models on code, recommending refactorings, and visualizing data in the IDE. We provide several examples of existing plugins that implement these cases. Finally, to make it easier to start working with the platform, we develop and provide simple plugins for each use case that could serve as a template for a new project.
\end{abstract}

\section{Introduction}\label{sec:introduction}

A lot of state-of-the-art approaches and tools in modern software engineering are data-driven, meaning that they rely on mining software repositories and analyzing the collected data. 
This leads to researchers and tool builders having to perform a lot of complex operations with the code, like building its various representations~\cite{alon2019code2vec}, processing the representations~\cite{falleri2014fine}, and using the results to power the data-driven techniques to assist with engineering tasks~\cite{dam2018deep}. Building the tools for code analysis and manipulation constitutes a significant part of the researchers' work. 

At the same time, Integrated Development Environments (IDEs) routinely perform similar processing tasks in the background during their use. This processing enables the modern IDEs to provide a rich set of capabilities for users, including code completion, automatic refactoring, and collaborative programming, making the development process more effective. 

JetBrains is one of the leading vendors of software engineering tools, including IDEs for various languages: IntelliJ IDEA for Java and Kotlin, PyCharm for Python, CLion for C and C++, and others. These IDEs are all based on the IntelliJ Platform.\footnote{The IntelliJ Platform: \url{https://www.jetbrains.com/opensource/idea/}} This is logical because, despite all the differences between languages, a lot of the functionality is reused between different IDEs. Some of this functionality includes the powerful features for working with the code mentioned above, so using the platform could save time for researchers when implementing their ideas.

Another important feature of the IntelliJ Platform is that it can be used to enhance IDEs by developing \textit{plugins}. Plugin developers have access to all of the platform APIs, and can therefore conveniently process code and other software artifacts. Moreover, when the plugin is ready, a developer can publish it in the Marketplace\footnote{Marketplace: \url{https://plugins.jetbrains.com/}} to make it available for millions of IDE users, thus opening up new possibilities for the evaluation of the idea.

In this paper, we briefly describe the IntelliJ Platform by listing its key features, such as its internal code representation, tools for interacting with the user's code, and the UI toolkit. Then, we list four potential use cases that software engineering researchers face, where the IntelliJ Platform can be employed: mining software data, running machine learning models, recommending refactorings, and visualizing information in the IDE. For each use case, we list several diverse examples of existing plugins and describe them.

Finally, to facilitate the use of the IntelliJ Platform, we create and share simple but complete template plugins for the same four use cases. These templates are intended to highlight the possibilities of the platform, and can be used as a base for more comprehensive plugins. All the examples are available online: \url{https://github.com/JetBrains-Research/refactoring-workshop-demo}.
\section{The IntelliJ Platform}\label{sec:platform}

The IntelliJ Platform provides many different features that can be useful to researchers and practitioners when building tools and implementing different approaches. 
In this section, we describe the main ones.

\subsection{Code Representation}
\begin{figure}[ht]
    \centering
        \parbox{.55\columnwidth}{
            \centering
            \subcaptionbox{An example code snippet.\label{fig:code_example}}{
                \includegraphics[width=\hsize]{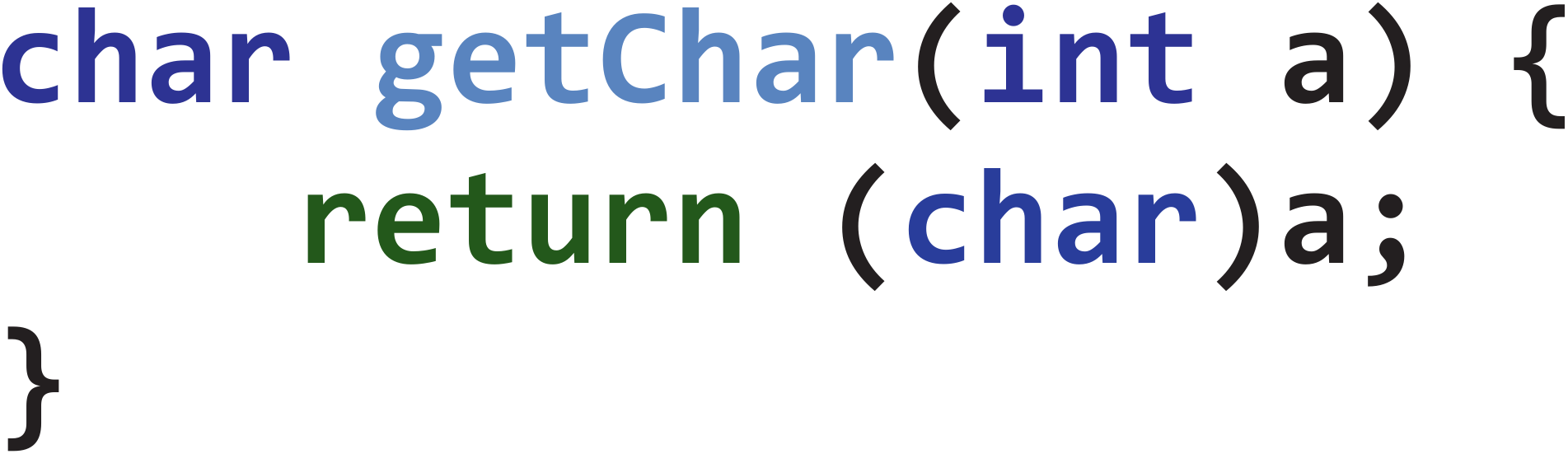}
            }
        }
        
        \vskip2em
        
        \parbox{.9\columnwidth}{%
            \subcaptionbox{The snippet's PSI tree. \texttt{PsiWhiteSpace} nodes are omitted for readability.\label{fig:psi_example}}{
                \includegraphics[width=\hsize]{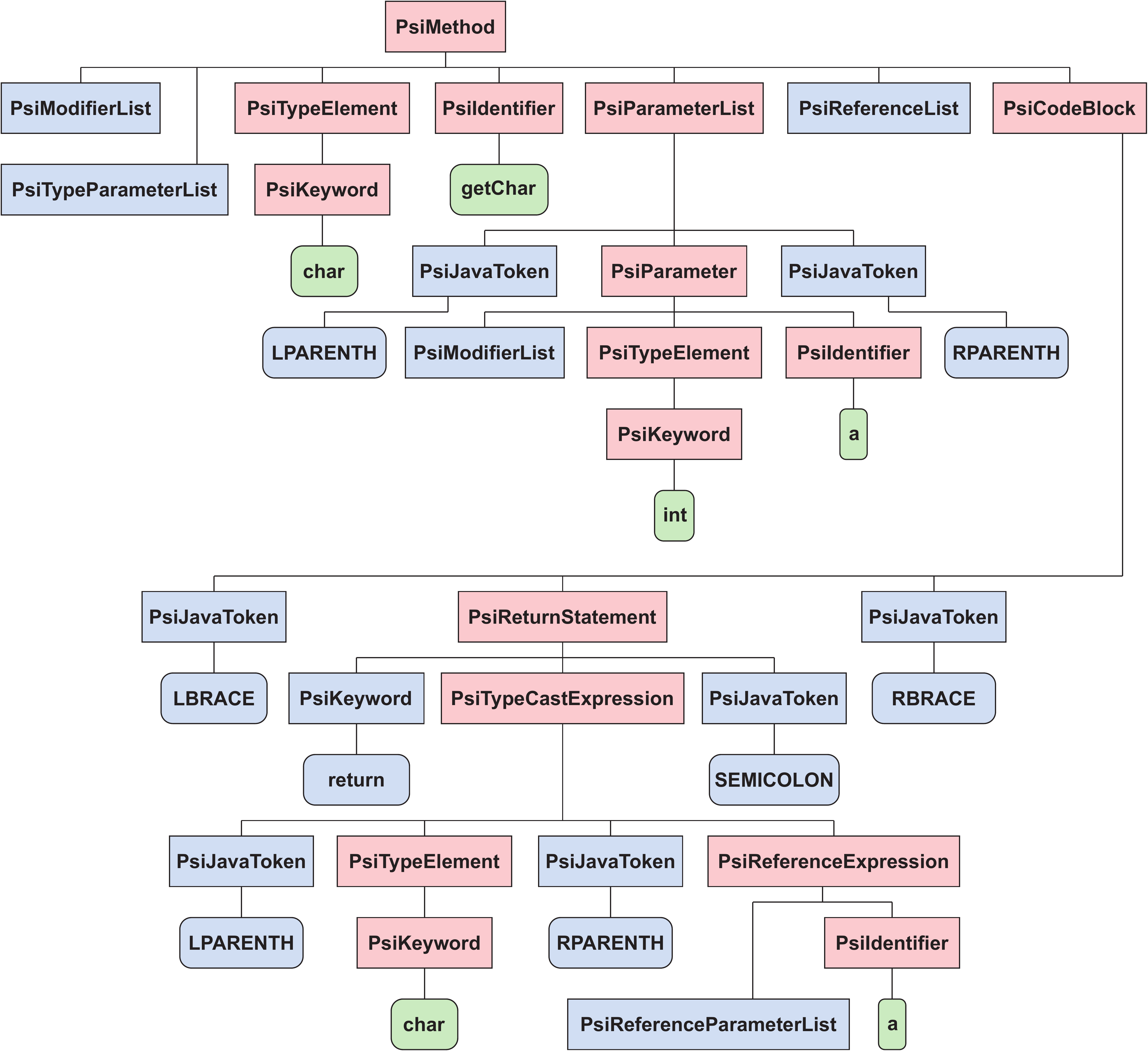}
            }
    }
\caption{An example of a code snippet with the corresponding PSI tree.}\label{fig:psi}
\vspace{-1em}
\end{figure}

The first stage of working with code in various research tasks is most often choosing a proper representation and converting the code into it. 
Researchers often use syntax trees to work with source code, since they capture the syntactic structure of code by design.
In the IntelliJ Platform, the \textit{Program Structure Interface (PSI)}\footnote{PSI: \url{https://plugins.jetbrains.com/docs/intellij/psi.html}} is responsible for parsing source code and building syntactic trees above it.
PSI trees are 
concrete syntax trees (CST), since they also contain whitespaces, punctuation, and can be used to infer some semantic information about the code.
\Cref{fig:psi} presents an example of a PSI tree for a small code snippet.

PSI allows the users of the API to extract a wide variety of things from the code: traverse elements of the exact type (\textit{e.g.}, methods, code comments, if-statements), method invocations, all statements from the particular code fragment, and so on. 
Other important features of PSI include the support for type inference, basic control-flow and data-flow analysis, as well as the ability to incrementally update the built representations. 

\subsection{Code Transformations}
The IntelliJ Platform allows to transform source code in several ways: from large scale changes like refactorings to smaller ones like fixing of typos or type casting, which are implemented as \textit{quick fixes}.
Modern IDEs support automatic refactorings such as Rename, Move, and Extract.
There is a set of refactorings already implemented in the IntelliJ Platform that could be triggered from plugins. This means, for example, that if a researcher develops a new approach to recommending a certain refactoring or to fixing a new code smell, they would most likely only need to implement the detection and the analysis in the plugin, not the low-level details of code transformation.

\subsection{Code Inspections and Intention Actions}
To raise developers' awareness of existing code issues, one can use \textit{code inspections}.
For example, one can implement a code inspection that checks the length of the code line and highlights if the length exceeds a certain limit.
A code inspection analyzes source code in the background mode, which means that it does not interrupt the user's workflow, and highlights the discovered problems as they are found.

To provide users with a possibility to trigger some code transformation upon request, one can implement \textit{Intention Actions}.
An intention action is an action that might be triggered by the use of some specific shortcut or by choosing the corresponding action in the list appearing after clicking a yellow bulb icon in the editor.

\subsection{User Interface Toolkit}
While developing a new tool, it is important to think about the way the users will interact with it. 
Naturally, it is more convenient for users when the tool is integrated into a familiar editor than to work with a stand-alone tool.
The IntelliJ Platform provides a great variety of ways to attract the user's attention to existing issues in the code.
One of them is highlighting the code lines that contain the problem. 
Another way is to send a notification to a user about the discovered problem. 

Taking action to fix the problem is also easier when the action is embedded into the editor.
When creating a plugin, developers have access to all the standard components of the editor UI, such as tool windows, popup menus, or buttons.
Depending on the nature of the plugin, it may use a simple UI, for example, a single tool window at the bottom of the editor with all the necessary information, or something more complex like inlaying information right into the code, displaying graphs or figures, or interacting with the writing process.

\section{Use Cases}\label{sec:use-cases}

To illustrate how the IntelliJ Platform can be helpful with building tools for specific software engineering tasks, in this section we will describe four different use cases and provide examples of their implementation in existing plugins.
We also collected a list of papers that describe different IntelliJ plugins: \url{https://zenodo.org/record/5443946}.

\subsection{Mining Data}

Mining software data is an essential step in research --- it can be used for large-scale code analysis, as well as for gathering quality data to train a machine learning (ML) model, for example, to detect code smells.
The IntelliJ Platform can help with mining data from code repositories. 
All information about the source code can be extracted and saved for further analysis and processing.

Moreover, using the IntelliJ Platform, researchers can not only analyse the source code itself, but also analyse the history of its changes. 
The platform provides the \textit{git4idea} module responsible for working with Git. 
It allows to extract commits from the given repository, traverse the changes, and analyse them. This functionality opens up a lot of possibilities, like analyzing the history of the user's code to make more personalised suggestions.

To use the platform for data mining, one can develop plugins that run in the \textit{headless mode}, which means that the UI of the IDE does not spin up.
This allows running plugins as command-line tools, for example, on a remote server. 
With such a setup, plugins can be configured to process multiple projects in one run. 
This gives researchers the ability to conveniently collect large datasets of diverse data from code.

An example of the IntelliJ-based plugin that was created to mine the data is PSIMiner~\cite{spirin2021psiminer}. This plugin allows researchers to mine PSI trees from projects and to create datasets for ML models from them. PSIMiner launches IntelliJ IDEA in the headless mode, processes the given projects, and saves the mined data into one of the several supported formats.

The data can be collected not only from a static project or its history, but also directly from the coding process. 
Cao et al.~\cite{cao2020tool} developed a plugin for IntelliJ IDEA and Android Studio called DevActRec that accurately and non-intrusively collects the user's activity during a programming session.
DevActRec collects information about the keystrokes, position of the cursor, dialogs, events in the IDE, etc. 
The plugin can also correlate this information with the higher-level activities, such as coding, debugging, navigating, etc.

\subsection{Running Machine Learning Models}

After an ML model is trained, it can be useful to embed it into the plugin to make it available to users. 
A lot of ML tasks in software engineering fit neatly into the IDE setting. Examples of such tasks are predicting the name of a method, generating documentation, suggesting a new class for a method, and discovering vulnerable code. 

ML models can be used in IntelliJ plugins through the \textit{KInference}\footnote{KInference: \url{https://github.com/JetBrains-Research/kinference}} framework for running ML models in the standardized ONNX\footnote{ONNX: \url{https://github.com/onnx/onnx}} format in Java and Kotlin applications. 
\emph{KInference} does not include the training functionality or use any native libraries, which allows for compact plugins. 
Low dependency size is a particularly important quality of \emph{KInference}: existing deep learning frameworks for JVM, such as \emph{deeplearning4j}\footnote{deeplearning4j: \url{https://deeplearning4j.org/}} and \emph{TensorFlow Java}\footnote{TensorFlow Java: \url{https://www.tensorflow.org/jvm/}} are bulky dependencies due to the heavyweight training functionality they include. 
Classical ML models can also be used in plugins through existing frameworks such as \emph{weka}.\footnote{weka: \url{https://www.cs.waikato.ac.nz/ml/weka/}}

An ML model is used in the Sorrel plugin~\cite{pogrebnoy2021sorrel}. 
Sorrel can detect licenses in the opened Java projects and help developers manage them and find incompatibilities. 
To detect licenses, the plugin uses an ML classifier that was exported in the ONNX format and processed in the IDE environment via KInference. 

Another example is $SWAN_{ASSIST}$~\cite{piskachev2019swan_assist}, a plugin for detecting security-relevant methods. 
This plugin uses a machine learning classifier to detect the necessary methods and allows the user to iteratively manually add new methods to the training dataset of the classifier, thus making the classifier better during its operation.
Similarly, ELFF~\cite{bowes2017getting} is an IntelliJ IDEA plugin that runs classifiers for defect prediction. 

\subsection{Recommending Refactorings}

Developing a new refactoring recommendation approach is a challenging task in itself, and the need to implement refactoring support makes it even harder. 
The IntelliJ Platform has refactorings that are already implemented and can be triggered from plugins.
For example, if one develops a new way to recommend Move Method refactoring, they only need to decide which method should be moved to which class and pass these values to the class of the platform responsible for the Move Method refactoring. 
All refactorings in the platform have a set of defined preconditions to ensure that the changes will not break the code.

An example of a plugin that detects and refactors code smells is cASpER~\cite{de2020casper}. cASpER can detect four different code smells (Feature Envy, Misplaced Class, Blob, and Promiscuous Package), visualize the results of the detection in a special window through radar maps, and then refactor the code if the user chooses to do so. 
The plugin relies on the IntelliJ Platform's APIs to carry out the refactoring itself.
 
DARTS~\cite{lambiase2020just} is a similar plugin, but developed for the detection and the refactoring of test smells.
It implements a mechanism to detect instances of three test smell types and enables their automated refactoring through the integrated APIs provided by the IntelliJ Platform.

\subsection{Visualizing Information}

Finally, sometimes it is of interest to show some information to the developer that might help them in their tasks. This can mean visualizing some statistics or focusing their attention on a specific part of code.

All of the above-mentioned plugins do this in one way or another, by adding tool windows or notifications. 
For example, Sorrel~\cite{pogrebnoy2021sorrel}, a plugin for license management, shows the licenses of dependencies declared in Gradle scripts by embedding the information straight into the editor.

For some plugins, visualization is the primary feature. 
For instance, RefactorInsight~\cite{kurbatova2021refactorinsight} supplements code diffs with information about performed refactorings. 
This allows the users to hide refactorings in commits and pull requests to focus on reviewing changes that alter the semantics of the code.

VITRuM~\cite{pecorelli2020vitrum} is a plugin that provides developers with an advanced visual interface of test-related metrics. 
This includes test coverage, test smells, and flakiness. 

If necessary, it is possible to equip an IntelliJ plugin with advanced graphical features. 
For example, DeepGraph~\cite{hu2018deepgraph} is a PyCharm plugin for visualizing and understanding deep neural networks. 
The plugin visualizes deep learning models in the IDE and uses JavaFX's WebEngine for visualization.
\section{Templates}

To help developers with creating future plugins, we prepared several basic plugin templates that cover the described use cases. 
These plugins are straightforward, showcase various sides of the IntelliJ Platform, and are intended to be used as a foundation for more advanced plugins. 
The source code and the pre-built versions of the plugins are available at \url{https://github.com/JetBrains-Research/refactoring-workshop-demo}.
In this section, we briefly describe the provided template plugins.

\subsection{Mining Data}

We implemented a plugin that collects Javadocs for methods as an example of a data mining task.
This plugin runs in the headless mode and can thus be used as a command-line tool.

The plugin receives two arguments: a path to the project that needs to be analyzed and a path to the output file. 
The plugin then launches IntelliJ IDEA in the background, detects and parses all Java files in the project, and extracts Javadocs for all methods where they are present. 
Finally, the methods and their Javadocs are saved to a JSON file.

This template can be easily modified to extract almost any other code entity by simply changing the analyzed PSI nodes.

\subsection{Running Machine Learning Models}

As an example of an ML-powered plugin, we chose to implement a significantly simplified version of the Sorrel~\cite{pogrebnoy2021sorrel} plugin for managing software licenses. 

The plugin inferences a model when a LICENSE file is opened in the editor. The classifier processes the text of the license and tries to recognize it. 
This simple model can recognize the three most popular permissive open source licenses: Apache-2.0, BSD-3-Clause, and MIT. 
If the license is recognized, the plugin shows a little message at the top of the editor with the name of the license.

The source code of this plugin can be used as an example of inferencing an ONNX model in a plugin.

\subsection{Recommending Refactorings}

To demonstrate the abilities of the IntelliJ Platform in carrying out refactorings, we built a very simple plugin that detects the Feature Envy smell and uses the Move Method refactoring to combat it.

In the given project, in the currently opened file, the plugin detects all classes within the file. Then the plugin traverses all the methods and checks if any of the methods have Feature Envy, meaning that the method accesses more fields of another class than the one that it belongs to.
If such methods are found, the plugin suggests the user to move them to the more appropriate class. If the user agrees, the plugin performs the Move Method refactoring automatically.

This template can be the foundation for almost any plugin for automated refactoring: it showcases the use of the IntelliJ Platform's refactoring APIs.

\subsection{Visualizing Information}

Finally, for visualization, we created a template of a plugin that collects and displays basic metrics for classes in the code.

Similarly to the previous plugins, this plugin traverses PSI trees of Java files to find all classes in the code. 
It then analyzes the PSI to calculate several basic metrics: number of fields, number of methods, and lines of code.
This information is displayed directly in the IDE through a dedicated tool window under the editor containing the metric values per class.

This plugin introduces the tool window --- a basic but important UI element, which can be used in almost any plugin to display the relevant information.

\section{Conclusion}\label{sec:conclusion}

In this paper, we showcase the IntelliJ Platform---a platform that IntelliJ-based IDEs run on---as a framework for researchers and developers to create various plugins and tools.
We list the main features of the platform, and then describe four example use cases that researchers might be interested in: mining software repositories, running ML models, recommending refactorings, and visualizing information in the IDE. 
We provide examples of existing IntelliJ plugins that fall under these use cases and demonstrate how the IntelliJ Platform can be useful for researchers. 
Finally, we develop and share basic plugin templates for each use case that can be used as a starting point for more complex plugins in the future.

We hope that the IntelliJ platform can serve as a bridge between state-of-the-art approaches and millions of developers worldwide. Examples for all use cases from this paper are available online: \url{https://github.com/JetBrains-Research/refactoring-workshop-demo}.

\bibliographystyle{IEEEtran}
\bibliography{IEEEabrv,paper}

\end{document}